\documentclass[]{article}
\usepackage{amsmath}
\usepackage {amssymb}
\usepackage{epsfig}
\renewcommand{\Im}{\mathop{\mathrm{Im}}\nolimits}
\renewcommand{\Re}{\mathop{\mathrm{Re}}\nolimits}
\renewcommand{\vec}[1]{\ensuremath{\mathrm{\mathbf{#1}}}}
\begin {document}
\title {Anomalous  absorption of light  by a nanoparticle and bistability in the presence of resonant fluorescent  atom}
\author {Gennady N. Nikolaev\\{\it Institute of  Automation and Electrometry of SB RAS}, \\ \it{ Pr. Koptyuga 1, Novosibirsk, 630090,
Russia}}
\date {November 16, 2004}
\maketitle
\begin{abstract}
Absorption of light by a nanoparticle in the presence of resonant
atom and fluorescence of the latter are theoretically
investigated. It is shown, that absorption of light by a
nanoparticle  can  be increased by several orders because of
presence of atom. It is established, that optical bistability in
such system  is possible.\\

\end{abstract}
\section{Introduction}\label{c:int}

 The cross-section of light adsorption by an isolated spherical nanoparticle imbedded in a host
medium and which radius {$a$} is essentially smaller then light
wavelength {$\lambda$} in the medium ($a/\lambda \ll 1$) is given
by the classical formula \cite{LanLiphVII}
\begin{equation}
\label{e0}
  {\sigma}_p=24 \pi \frac{a}{\lambda}\frac{\epsilon ''}{|\epsilon +2 |^2}S\, ,
\end{equation}
where ${S}=\pi a^2$, ${\epsilon}\equiv
\varepsilon_{p}/\varepsilon_{h}=\epsilon '+i\epsilon ''$ is the
relative complex dielectric function of the nanoparticle,
$\varepsilon_{h}$ and $\varepsilon_{p}$ are dielectric functions
of the nonabsorbing host medium and nanoparticle respectively.

As a rule, ${\sigma}_p$ is smaller then geometrical cross-section
of the nanoparticle $S$. On the other hand, it is well known that
the cross-section of resonant atom-light interaction is
considerably larger:
\begin{equation}
\label{e0a}
  {\sigma}_a =\frac{1}{2\pi}\frac{\mathstrut{\gamma_{0h}}}{\gamma}{\lambda}^{2} \, ,
\end{equation}
where $\gamma_{0h}$ and {$\gamma$} are the radiation and the total
width of the resonant transition of an atom imbedded in a host
medium. Note that $\gamma_{0h}$ is expressed in terms of
free-space spontaneous emission rate $2\gamma_{0}$ as $\gamma_{0h}
\equiv (\varepsilon_{h})^{1/2}|(\varepsilon_{h}+2)/3|\gamma_{0}$
(see, e.g., \cite{BaHuLo92, BaHuLo96}).

\begin{figure}[htbp]
\noindent
\begin{center}
\epsfig{figure=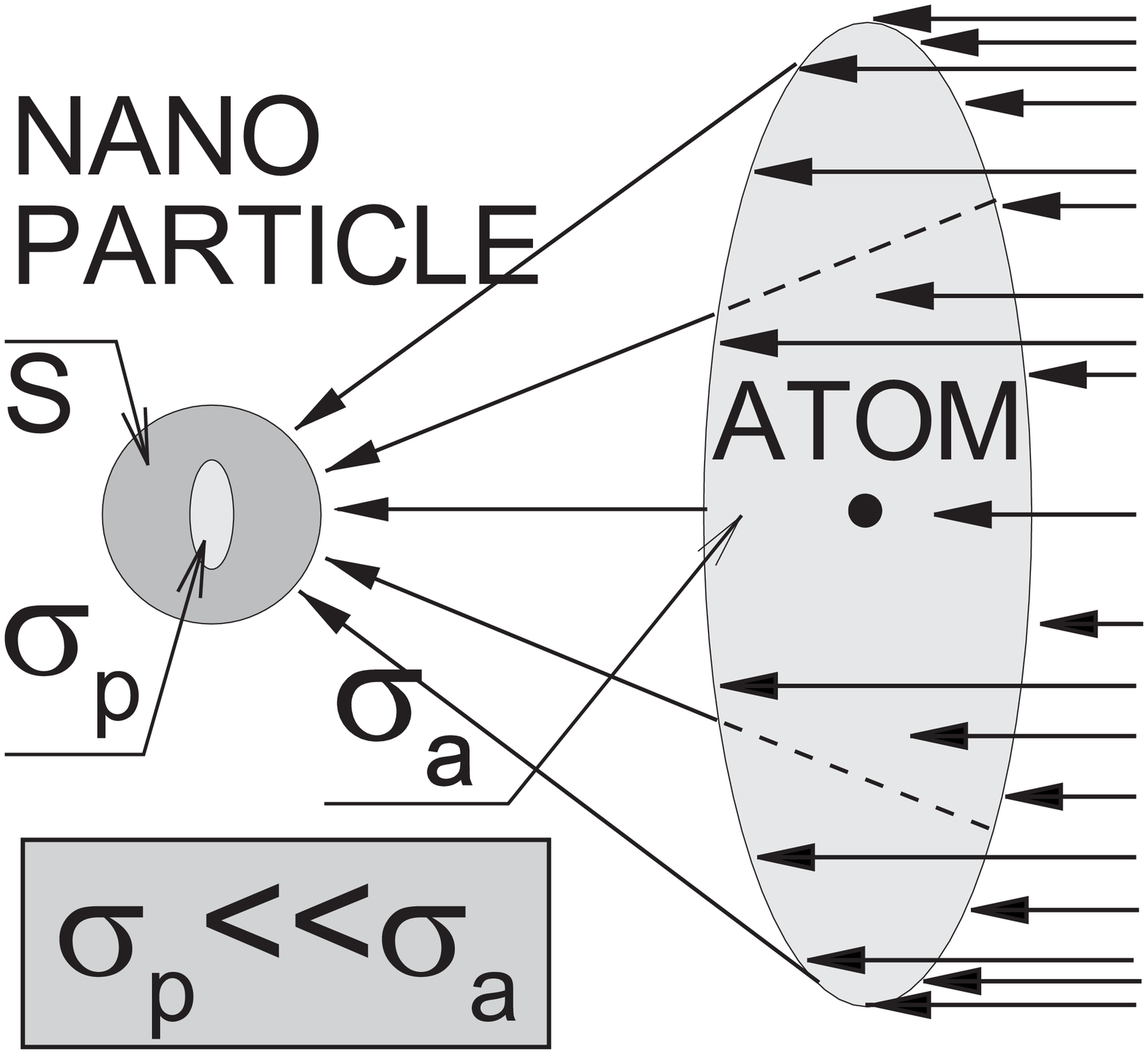,width=0.6\linewidth}
\end{center}
\caption{Atomic lens.}
\label{fig1}
\end{figure}

As a rule, ${\sigma}_p \ll {\sigma}_{a}$ (see Fig.~\ref{fig1}).

The aim of the paper is clarification of the probability of
cascade energy transfer from light to an atom and then to a
nanoparticle \cite{Nik89}, and investigation of atomic
fluorescence in this conditions.
\section{Oscillating classical dipole}\label{c:cl_dip}
Let us consider an auxiliary problem connected with atomic
excitation transfer to a nanoparticle, the particle absorption of
the electric field energy of classical dipole ${\vec{d}} \equiv
d\{\sin(\psi),0,\cos(\psi)\}$ which oscillates with frequency
{$\omega$} and is at a distance {$\vec{R}$} from the nanoparticle
center. The power {$Q_c$} absorbed by the nanoparticle can be
represented in the form
\begin{equation}
\label{e1b}
  Q_{c}=-\overline{\dot{\vec{d}}(\vec{R}) \, \delta \vec{E}(\vec{R})} \,,
\end{equation}
where {$\delta \vec{E}(\vec{R})$} is the `image field' of the
dipole, and overline denotes time averaging over the time that is
considerably greater than period of light wave. Since $\delta
\vec{E}$ depends linearly on $\vec{d}$, the absorbed power {$Q_c$}
can be rewritten as
\begin{equation}
\label{e2}
   Q_{c}=2\omega \Im \left[\sum_{\alpha, \beta} G_{\alpha\beta}(\vec{R}, \vec{R};\omega)
  {\vec{d}}_{\beta}{\vec{d}}_{\alpha} \right] \, ,
\end{equation}
where {$G_{\alpha\beta}(\vec{R}, \vec{R};\omega)$} is the field
susceptibility (or tensor-valued Green function):
\begin{equation}
\label{e3}
\delta\vec{E}_{\alpha}(\vec{r},\omega)=\sum_{\beta}G_{\alpha\beta}(\vec{r},
\vec{r}';\omega){\vec{d}}_{\beta}(\vec{r}',\omega)\, .
\end{equation}
\subsection{Field susceptibility}\label{c:F_Susc}
In the near zone ($R\ll \lambda$) the `image field' of the dipole
(and, consequently, $G$) can be found by solving electrostatic
Laplace equation. One can start with the scalar potential of a
single charge $e$ located on the axis $Z$ at the distance $R$ from
the center of the particle \cite{Sratt41}):
\begin{equation}
\label{e4a}
   \varphi_{e}(\vec{r,R})=-\frac{e}{a}\sum_{n=0}^{\infty}{\frac{(\epsilon -1)n}{(\epsilon
   +1)n+1}\,P_{n}\left(\cos(\theta)\right)\left(\frac{a^2}{rR}\right)^{n+1}},
\end{equation}
where $P_{n}(\cos(\theta))$ is the Legendre polynom and $\theta$
is an elevation angle of the vector $\vec{r}$.

The potential of the point-like dipole $\vec{d}$ located on the
axis $Z$ at the distance $R$ from the center of the particle is
the sum of potentials (\ref{e4a}) caused to nearly situated
charges $e$ and $-e$
\begin{eqnarray}
\label{e4b}
   &&\varphi(\vec{r,R})=\frac{d}{aR}\sum_{n=0}^{\infty}\frac{(\epsilon
   -1)n}{(\epsilon +1)n+1}\,\bigg [(n+1)\cos(\psi)P_{n}\left(\cos(\theta)\right)\nonumber \\
   &&\phantom{\varphi(\vec{r,R})}\left.-\sin(\psi)\sin(\theta)
   \cos(\phi)\frac{dP_{n}\left(\cos(\theta)\right)}{d\cos(\theta)}\right]\left(\frac{a^2}{rR}\right)^{n+1} ,
\end{eqnarray}
where $\phi$ is an azimuth angle of the vector $\vec{r}$.

 So, the `image field' and $G$ are found from (\ref{e4b}) and represented as a series (see, for
 example, \cite{KliDuLe96}). This series diverges when $R \to a$, so that
the higher terms start play the major part in it. Fortunately, one
possible to rewrite it in a reasonable way, so that the field
susceptibility tensor can be expressed in the form
\begin{eqnarray}
\label{e4}
   G_{zz}(\vec{R})={\frac{\epsilon -1}{\epsilon
   +1}\,\frac{\mu^{3}}{\varepsilon_{h}R^3}}
   \left[{\frac{2}{{\varrho}^3} } {\,+\frac{1}{{\varrho}^2{\rho}}
   +\frac{1}{{\varrho}{\rho}^2}} {-\frac{{\epsilon}^{-1}}{{\rho}^{3}}\,
  \frac{\epsilon+1}{\epsilon+2}\,F\left(\begin{array} {c}{1,1+\nu}\\ {2+\nu}\end{array}\Biggr |
  {\rho}^{-2}\right)}\right], &&\\
  G_{xx}(\vec{R})={\frac{1}{2}\,\frac{\epsilon -1}{\epsilon +1}\,\frac{\mu^{3}}
  {\varepsilon_{h}R^{3}}} \left[{\frac{2}{{\varrho}^3}}{\,- \frac{{\epsilon}^{-1}}
  {{\varrho}^2{\rho}}-\frac{{\epsilon}^{-1}}{\varrho{\rho}^2}} {+\frac{{\epsilon}^{-2}}{{\rho}^{3}}\,
  \frac{\epsilon+1}{\epsilon +2}\,F\left( \begin{array}{c}{1,1+\nu}\\{2+\nu}
  \end{array}\Biggr | {\rho}^{-2}\right)}\right],&&
\label{e5}
\end{eqnarray}
where ${\mu} \equiv \epsilon/(\epsilon+1)$, ${\varrho} \equiv
{\mu}{\varrho}_{0}^{}$, ${{\varrho}_{0}^{}} \equiv
({\rho}-{\rho}^{-1})$, ${\rho} \equiv R/a$, ${\nu} \equiv
1/(\epsilon +1)$, ${\epsilon} \equiv
\varepsilon_{p}/\varepsilon_{h}$, {$\varepsilon_{p}$} and
{$\varepsilon_{h}$} are complex dielectric function of the
nanoparticle and host surroundings accordingly,
{$F\left(\begin{array}{c}{a,b}\\c\end{array}\Biggr | s\right)$} is
hypergeometric function.

\begin{figure}[htbp]
\noindent
\begin{center}
\epsfig{figure=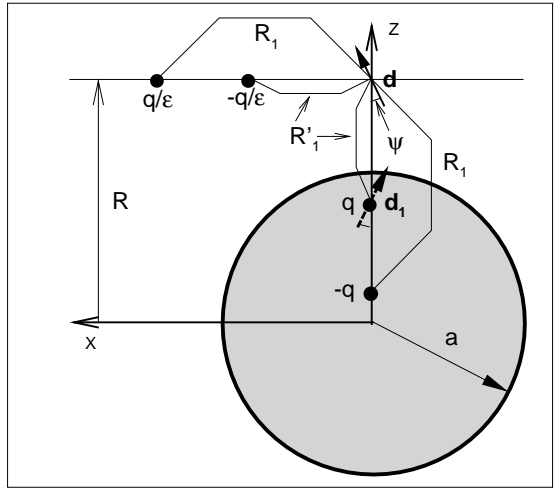,width=0.6\linewidth}
\end{center}
\caption{Geometry of the problem.}
\label{fig2}
\end{figure}

Let us concider some {\bf limit cases}.
\begin{itemize}
    \item \underline{$|\epsilon |\gg 1$} (i.e., ideal conductor)
        \begin{eqnarray}
        \label{e6}
            G_{zz}(\vec{R})&\simeq&{\frac{1}{\varepsilon_{h} R^{3}}}
            \left[{\frac{2}{{\varrho}_{0}^3}}{\,+\frac{1}{{\varrho}_{0}^2{\rho}}+
            \frac{1}{{\varrho}_{0}^{}{\rho}^2}} \right], \\
            G_{xx}(\vec{R})&\simeq&{\frac{1}{\varepsilon_{h}R^{3}{\varrho}_{0}^3}}\, .
        \label{e7}
        \end{eqnarray}
As is well known, in this case the `image field' that corresponds
to (\ref{e6}) and (\ref{e7}) can be represented by the sum of the
fields of the charge ${q_{0}} \equiv ad_{z}/R^2$ and $-q_{0}$, and
of the dipole ${\vec{d}}_{0} \equiv (a/R)^3
d\{-\sin(\psi),0,\cos(\psi)\}$ field. The charge -${q_{0}}$ is
placed at the particle center; the charge ${q_{0}}$ and dipole
${\vec{d}}_{0}$ are located along $Z$ axis at the distance
$a{\varrho}_{0}^{}$ from dipole $\vec{d}$(another words, at
distance $a^2/R$ from the particle center toward the dipole
$\vec{d}$).
\end{itemize}
Similarly to this limit case, the first three terms in square
brackets of the  equations (\ref{e4}) and (\ref{e5}) can be
interpreted as the sum of the fields of the charges ${q} \equiv
\mu^{3} (\epsilon-1)/(\epsilon+1)q_{0}$, $-q$, ${q_{\epsilon}}
\equiv -q/\epsilon$, $-q_{\epsilon}$, and of the dipole
${{\vec{d}}_{1}} \equiv \mu^{3}
(\epsilon-1)/(\epsilon+1){\vec{d}}_{0}$. Charge ${q}$ and dipole
${{\vec{d}}_{1}}$ are located along $Z$ axis at the distance
$R'_{1}\equiv a{\varrho}$ from dipole $\vec{d}$, whereas the
charge $-q$ is disposed at the distance $R_{1}\equiv a{\rho}
\left[\varrho/(\rho/\mu-{\rho}^{-1})\right]^{1/2}$ from dipole
$\vec{d}$  along the same $Z$ axis. Charges ${q_{\epsilon}}$ and
${-q_{\epsilon}}$ are located at the distance $R_{1}$ and $R'_{1}$
respectively from dipole $\vec{d}$ along $X$ axis (see Fig.~\ref{fig2}).
 These terms of the  equations (\ref{e4}) and (\ref{e5}) dominate while as $R/a \gtrsim 1$, and $\epsilon$ is far away from
the region of the surface multipole resonances that take place
when $\Re(n+1+\nu) = 0$.
\begin{itemize}
    \item Small distances, \underline{$R/a-1 \ll 1$ ($\varrho_{0}\simeq
    0$)}
    \begin{equation}
    \label{e7s}
        G_{xx}(\vec{R})\simeq\frac{1}{2}\,G_{zz}(\vec{R})\simeq\frac{\epsilon-1}{\epsilon +1}
        \,\frac{\mu^{3}}{\varepsilon_{h}R^{3}}\,\frac{1}{{\varrho}^3}.
    \end{equation}
    It is exactly the case of the planar interface.
    \item Large distances, \underline{$R/a \gg 1$ (i.e., $\varrho_0\simeq
    R/a$)}
    \begin{equation}
    \label{e8}
        G_{xx}(\vec{R})\simeq\frac{1}{4}\,G_{zz}(\vec{R})\simeq\frac{\epsilon -1}{\epsilon+2}
        \,\frac{a^{3}}{{\varepsilon}_{h}R^6}\,.
    \end{equation}
    The `image field' corresponding to (\ref{e8}) is given by
    \begin{equation}\label{e9}
        \delta \vec{E}(\vec{R})\simeq\frac{\epsilon -1}{\epsilon+2}\,
        \frac{a^{3}}{{\varepsilon}_{h}R^6}\left[3(\vec{n}\vec{d}\,)\vec{n}+\vec{d}\,\right]\,,
    \end{equation}
    where $\vec{n}$ is the unit vector in the direction from the
    center of the particle to dipole $\vec{d}$.
\end{itemize}
\textbf{Physical interpretation of the formula (\ref{e9})}. At
large distance the electric field ${\vec{E}_{\vec{d}}}$ of the
dipole $\vec{d}$ is homogeneous in the vicinity of the particle.
The field of the polarization (or scattered field) $\delta\vec{E}$
of the particle in such homogeneous field $\vec{E}_{\vec{d}}$ at
the location of the dipole $\vec{d}$ is given by (see, for
example, \cite{LanLiphVII})
\begin{equation}
\label{e10}
   \delta \vec{E}(\vec{R})=\frac{\epsilon -1}{\epsilon+2}\,\frac{a^{3}}{R^3}
    \left[ 3(\vec{n} \vec{E}_{\vec{d}})\vec{n}-\vec{E}_{\vec{d}} \right] \, .
\end{equation}
In turn, the field $\vec{E}_{\vec{d}}$ in the quasistatic
approximation is given, as is well known, by \cite{LanLiphVII}
\begin{equation}
\label{e11}
   \vec{E}_{\vec{d}}=\frac{1}{{\varepsilon}_{h} R^3}
    \left[ 3(\vec{n}\vec{d})\vec{n}-\vec{d}\right] \, .
\end{equation}
Substituting this expression in (\ref{e10}) we get (\ref{e9}).
\section{Quantum consideration}\label{c:qu_dip}
Energy transfer to the nanoparticle from the real atom exited by
the light is given by (instead of (\ref {e2}))
\begin{equation}
\label{e11q}
  Q_{c}=2\omega \Im \left[\sum_{\alpha, \beta} G_{\alpha\beta}( \vec{R};\omega)
  \langle \colon\hat{\vec{d}}_{\beta}^{-}\hat{\vec{d}}_{\alpha}^{+}:\rangle\right]\,,
\end{equation}
where symbols `$\colon \, \colon$' and `$\langle \cdots\rangle$'
denote the normal ordering operator and the quantum averaging
respectively. For two-level atom $\hat{\vec{d}}^{+}$ is
\begin{equation}
\label{e12}
  \hat{\vec{d}}^{+}=\left \{\hat{\vec{d}}^{-}\right\}^{\dag}=
  \vec{d}_{nm}\exp\left(i\Omega_{0}t\right)\hat{\sigma}_{-}\,,
\end{equation}
where  $\Omega_{0}\equiv \omega-\omega_{mn}$, $\vec{d}_{mn}$ and
$\omega_{mn}$ are matrix element of the dipole moment of the $m-n$
atomic transition and the resonance frequency of this transition,
$\hat{\sigma}_{\pm}$ are the raising and lowering Pauli's
operators.

Substituting (\ref {e12}) in (\ref {e11q}) results in
\begin{equation}
\label{e13}
  Q_{c}=2\omega \Im \left[\sum_{\alpha, \beta} G_{\alpha\beta}(\vec{R};\omega)
  {\vec{d}_{nm}}_{\beta}{\vec{d}_{mn}}_{\alpha}
  \right]\rho_{m} \equiv 2\hbar\omega \gamma_{c} \rho_{m}\, ,
\end{equation}
where $\rho_{m}$ is population of the upper atomic level $m$, and
$\gamma_{c}$ is the addition nonradiative broadening of the atomic
transition due to energy transfer from the atom to the particle
(see below).
\section{Density matrix}\label{c:dens_mat}
Density matrix for two-level atom in the vicinity of the
nanoparticle obey the follow system of the equations \cite{Nik90}
\begin{eqnarray}
\label{e14a}
  \frac{d\rho}{dt}&=&-(\gamma -i\Omega)\rho -i{\Omega}_{R}\Delta /2 \, , \\
  \frac{d\Delta}{dt}&=&-2\gamma(1+\Delta)- 2\Im \left[\Omega_{R}{\rho}^{*}\right] \, .
  \label{e14b}
\end{eqnarray}
where ${\Delta} \equiv \rho_{m}- \rho_{n}=\langle {\hat \sigma
}_{3} \rangle$ and ${\rho} \equiv \langle {\hat \sigma }_{-}
\rangle \exp(i\omega t)$ are the population difference and
coherence of combining levels,
${\Omega_{R}}=|\vec{E}\cdot\vec{d}_{mn}|/\hbar$ is Rabi frequency,
$\vec{E}$ is an local electric field acting  on the atom,
${\Omega}=\Omega_0 +\nu$, ${\nu} =\Re (\Gamma)$, ${\gamma} =\Im
(\Gamma)$,
\begin{equation}
\label{e15}
  \Gamma = \frac{1}{\hbar}\, \left[\sum_{\alpha,\beta} (\vec{d}_{mn})_{\alpha}
  G_{\alpha \beta}^{ex}(\vec{R})(\vec{d}_{mn}^{*})_{\beta}\right]\,.
\end{equation}
Here $G^{ex}$ is the exact field susceptibility. Therefore the
imaginary part of Eq. (\ref{e15}) as the total decay rate,
$\gamma=\gamma_{r}+\gamma_{c}$, describes both radiative decay
$\gamma_{r}$ and nonradiative one $\gamma_{c}$.

The field susceptibility, $G$, Eqs. (\ref{e4})--(\ref{e5}), is the
quasistatic approximation of the exact one, $G^{ex}$. Hence, it is
responsible for nonradiative decay only.  Nevertherless, the
radiative part of the decay rate can be found in our case by use
of quasistatic solution (\ref {e4b}). Indeed, the solution
represents multipole expansion of the scalar potential induced by
dipole $\vec{d}$. In the near zone the dipole part of the induced
potential (\ref {e4b}) is described by its term with $n=1$
\begin{equation}
\label{e16a}
    \varphi^{(1)}(\vec{r,R})=\frac{d}{r^2}\,\frac{\epsilon
   -1}{\epsilon +2}\,\frac{a^3}{R^3}\,\left[2\cos(\psi)\cos(\theta)
   -\sin(\psi)\sin(\theta)\cos(\phi)\right]\equiv\frac{{\vec{d}}_{p}\cdot\vec{r}}{r^3}\,,
\end{equation}
\begin{equation} \label{e16b}
    {\vec{d}}_{p}\equiv d \,\frac{\epsilon -1}{\epsilon
    +2}\,\frac{a^3}{R^3} \,\{-\sin(\psi), 0, 2\cos(\psi)\}.
\end{equation}
Dipoles $\vec{d}$ and ${\vec{d}}_{p}$ oscillate in phase due to
the inequality $a, R \ll \lambda$. So, the emission probability
and intensity of radiation are proportional to the total dipole
squared $|\vec{d}$ + ${\vec{d}}_{p}|^2$. Hence, the radiative part
of the spontaneous decay rate of an  atom placed next to a
nanoparticle is given by
\begin{equation}
\label{e16c}
    {\gamma_r} ={\gamma_{0h}}\,\left[\left|1-\frac{\epsilon -1}{\epsilon
    +2}\,\frac{a^3}{R^3}\right|^2 \sin^2(\psi)+\left|1+2\frac{\epsilon -1}{\epsilon
    +2}\,\frac{a^3}{R^3}\right|^2 \cos^2(\psi)\right],
\end{equation}
Expression (\ref{e16c}) agrees with the radiative part of the
spontaneous decay rate calculated without assumption of the
quasistatic approximation in the limit $a/\lambda \to 0$
\cite{KliDuLe01}.

 Steady-state solution of the equations
(\ref{e14a}), (\ref{e14b}) is conveniently expressed by
\begin{equation}
\label{e16}
  \rho_{m}=\frac{1}{2}\, \frac{{\tilde{\gamma}}^{-2}\xi^2 g(\tilde{\Omega})\tilde{I}/2}
  {1+\mathstrut {\tilde{\gamma}}^{-2}\xi^2 g(\tilde{\Omega})\tilde{I}/2} \,  ,
\end{equation}
where following dimensionless quantities are introduced: the total
broadening of the transition ${\tilde{\gamma}} \equiv
{\gamma_{r}}/{\gamma_{0h}}+{\gamma_{c}}/{\gamma_{0h}}$
\begin{eqnarray}
\label{e17}
  \tilde{\gamma}&\equiv& \left[\left|1-\frac{\epsilon -1}{\epsilon
    +2}\,\frac{a^3}{R^3}\right|^2 \sin^2(\psi)+\left|1+2\frac{\epsilon -1}{\epsilon
    +2}\,\frac{a^3}{R^3}\right|^2 \cos^2(\psi)\right] \nonumber\\
    &+& k^{-3}\Im \left[G_{zz}(\vec{R})\cos^2(\psi)+G_{xx}(\vec{R})\sin^2(\psi)\right] 
\end{eqnarray}
($k$ is the wave vector of the light in the medium, ${\psi}$ is
the angle between $\vec{R}$ and $\vec{d}\sim\vec{E}$); local field
gain factor ${\xi} \equiv \left|{E}/{E_0}\right|$
\begin{equation}
\label{e18}
  \xi =\left|\left\{\vec{e}_0+{\frac{\epsilon -1}{\epsilon+2}
  \frac{a^{3}}{R^3}\left[3(\vec{e}_0\cdot\vec{n})\vec{n}-\vec{e}_0\right]}\right\}\right| 
\end{equation}
(${E_0}$ and ${\vec{e}_0}$ are amplitude and unit polarization
vector of the incident light wave); formfactor of the optical
transition line ${g(\tilde{\Omega})}/\pi\equiv
\pi^{-1}/(1+{\tilde{\Omega}}^2)$;
${\tilde{\Omega}}\equiv\Omega/(\gamma_r+\gamma_{c})$; and
intensity of the incident light ${\tilde{I}}\equiv I/I_s$, where
${I_s}\equiv\hbar\omega\gamma_{0h}/\sigma_{a}$ is saturated
intensity of free atom in the medium, ${\sigma_{a}}$ is resonance
cross-section (\ref{e0a}). 
 The second term in the braces of (\ref{e18}) multiplied by $E_0$
is the scattered field of the nanoparticle in the near zone.
\section{Efficiency of the cascade energy transfer}\label{c:effi}
Efficiency of the cascade energy transfer ${\eta} \equiv Q_c/Q_p$
(where ${Q_p}=I\sigma_p$) can be found by using expressions
(\ref{e0}), (\ref{e13}) and (\ref{e16})
\begin{equation}
\label{e19}
  \eta\equiv\frac{1}{24\pi}\, \frac{\lambda}{a}\, \frac{\sigma_{a}}{S}\,
  \frac{|\epsilon +2|^2}{\epsilon ''
  I}\,\frac{\frac{1}{2}\tilde{\gamma}_{c}
  {\tilde{\gamma}}^{-2}\xi^2 g(\tilde{\Omega})\tilde{I}}
  {1+\frac{1}{2}\mathstrut\tilde{\gamma}^{-2}\xi^{2} g(\tilde{\Omega})\tilde{I}}.
\end{equation}

Obvious, efficiency of the cascade energy transfer can rich as
much as several orders of magnitude because of multiplication of
large values $\lambda/a$, $\sigma_{a}/S$, $|\epsilon
+2|^2/\epsilon ''$.
\subsection*{The limit cases}
\begin{itemize}
   \item Large distances, {\underline{$R/a \gg 1$}}\\
        In this case $\tilde{\gamma} \to 1$,  $\tilde{\gamma}_{c}
        \sim R^{-6}$ (see Eqs. (\ref{e17}) and (\ref{e8})). Hence, Eq. (\ref{e19}) is became
        \begin{equation*}
           \eta \sim R^{-6}.
        \end{equation*}
    \item Small distances, {\underline{$R/a-1 \ll 1$}}\\
        In this conditions $\tilde{\gamma}\approx\tilde{\gamma}_{c}\sim (R/a-a/R)^{-3}$
        (see Eqs. (\ref{e17}) and (\ref{e7})). Assuming in addition
        $\frac{1}{2}\mathstrut\tilde{\gamma}^{-2}\xi^{2}
        g(\tilde{\Omega})\tilde{I}\ll 1 $, we get from Eq. (\ref{e19})
        \begin{equation*}
        \eta\sim(R/a-a/R)^{3}.
        \end{equation*}
This decreasing of $\eta$ is rather unexpected because  the
probability of the energy transfer from atom to nanoparticle is
about 1 in this case. However, atomic resonance cross-section is
decreased when  $\gamma$ is increased (see Eq.(\ref{e0a})).
\end{itemize}
\begin{figure}[htbp]
\noindent
\begin{center}
\epsfig{figure=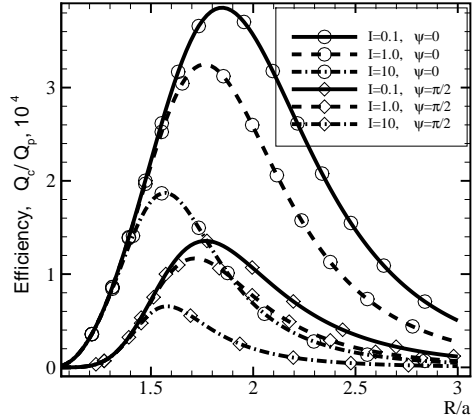,width=0.6\linewidth}
\end{center}
\caption{Efficiency of the cascade energy transfer as a function
of the dimensionless distance $R/a$ between resonance atom and
centre of the nanoparticle. It is supposed that the light
frequency is tuned in to the resonance atomic frequency at any
distance. Calculations are made for a silver nanoparticle
($\varepsilon_p =-15.37+i 0.231$, $\lambda =6328$ nm)}
\label{fig3a}
\end{figure}

So, the efficiency is decreased both for  large and small
distances between atom and particle. Therefore, it achieves a
maximum at an intermediate $R\gtrsim a$. Fig.~\ref{fig3a}  shows
efficiency of the cascade energy transfer $\eta$ versus $R/a$ in
the assumption of the exact resonance at any $R/a$.
\begin{figure}[hbtp]
\noindent
\begin{center}
\epsfig{figure=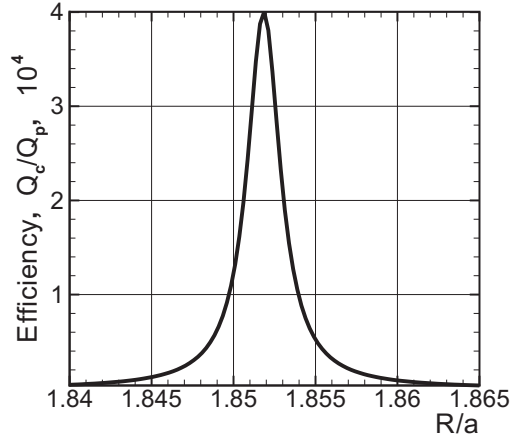,width=0.6\linewidth}
\end{center}
\caption{The same as in Fig.~\ref{fig3a}, but for fixed light
frequency $\tilde{\Omega}\equiv\Omega/\gamma_0=-410$.}
\label{fig3b}
\end{figure}
On the other hand, figure Fig.~\ref{fig3b} demonstrates very sharp
resonance dependence of $\eta$ as a function of $R/a$ when the
frequency of the light wave is fixed.

This sharply outlined resonance  can be used to determine location
of an  atom regarding the surface with subnanometer precision.
\section{Fluorescence}\label{c:flu}
As it is known, the intensity of fluorescence $I_f$ is
proportional to $\gamma_{r}\rho_{m}$. Using Eqs.
(\ref{e16c})--(\ref{e16}), intensity of fluorescence can be
written by
\begin{equation}
\label{e19f}
  I_f=I_0\,\frac{\frac{1}{2}\tilde{\gamma}_{r}
  {\tilde{\gamma}}^{-2}\xi^2 g(\tilde{\Omega})\tilde{I}}
  {1+\frac{1}{2}\mathstrut\tilde{\gamma}^{-2}\xi^{2} g(\tilde{\Omega})\tilde{I}}.
\end{equation}
\begin{figure}[thbp]
\noindent
\begin{center}
\epsfig{figure=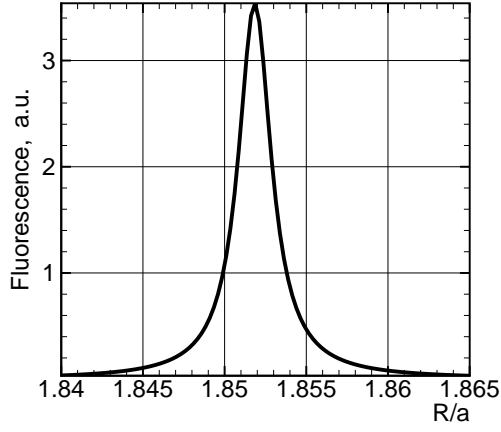,width=0.6\linewidth}
\end{center}
\caption{Intensity of fluorescence, a.u., as a function of $R/a$.
Conditions are the same as in Fig.~\ref{fig3b}.}
\label{fig4}
\end{figure}
Fig.~\ref{fig4} shows the intensity of fluorescence, a.u., as a
function of $R/a$. Conditions are the same as in Fig.~\ref{fig3b}.
\section{Heating of the particle}\label{c:heat}
\subsection*{Basic approaches and approximations}
\begin{itemize}
    \item Steady-state approximation
    \item Uniform temperature ${T_p}$ inside the particle
    \item Laplace's equation $\Delta T_h =0$ for the temperature of the host medium
    \item Energy balance equation\\ $Q_{c} +Q_{p} =
    \varkappa\int\vec{\nabla}T_{h}\,d\vec{S}$\\where ${\varkappa}$
    is thermal conductivity of surroundings
    \item Linear temperature dependence of $\varepsilon_{p} ''$:\\ $\varepsilon_{p}
    ''=\varepsilon_{p0} ''+{\alpha} (T_p-T_0)$
    \item $|\varepsilon_{p} ''/(\varepsilon_{p}'+\varepsilon_{h}')|$,
    $|\varepsilon_{h}/\varepsilon_{p}'|\ll 1$
\end{itemize}

Solution of the Laplace's equations is
\begin{equation}
\label{e20}
    T_h(r)=T_0+\frac{a}{r}(T_{p}-T_{0})\,.
\end{equation}
It gives a linear temperature dependence of heat removing from the
particle
\begin{equation}
\label{e21}
    Q_T=Q_{c} +Q_{p}=4\pi\varkappa a(T_{p}-T_{0})\,.
\end{equation}
Substituting (\ref{e0}) and (\ref{e19}) in this equation results
in following cubic equation with respect to the relative increase
of the image part of the dielectric function of the nanoparticle
${z}\equiv \delta\varepsilon_{p}''/\varepsilon_{p0}''$:
\begin{equation}
\label{e22}
    (x-y-1)z^3+\left[3(x-y)-2]+[3(x-y)-1+(x-1)f\right]z+xf+x-y=0\,,
\end{equation}
where the following dimensionless quantities are introduced:
${f}\equiv \frac{1}{2}\mathstrut\tilde{\gamma}^{-2}\xi^{2}
g(\tilde{\Omega})\tilde{I}$ is saturation factor, ${x}\equiv
(N_{p}+1)z/N$, ${y}\equiv \Delta z/N$, ${z/N} \equiv \left[{\hbar
\omega_{mn} \gamma_{c}}/({{4\pi a \varkappa}})\right]
\left[{\alpha}/{\varepsilon_{p0}''}\right]$, ${N_{p0}}\equiv
Q_{p0}/(\hbar \omega_{mn} \gamma_{c})$ and ${N}\equiv Q_{T}/(\hbar
\omega_{mn} \gamma_{c})$ are the number of photons absorbed by the
nanoparticle during the time $\gamma_{c}^{-1}$ directly from the
light wave and the total one respectively.

As well known, the qubic equation (\ref{e22}) may have 3 solution
at some parameters. Therefore, $z$ may exhibit bistable behaviour.
In the Fig.~6 it is shown the regions of such bistability.
\begin{figure}[htbp]
\noindent
\begin{center}
\epsfig{figure=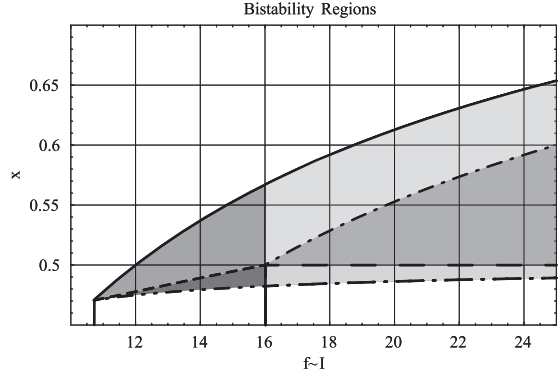,width=0.6\linewidth}
\caption{Bistability regions of the the relative increase of the
image part of the dielectric function of the nanoparticle ${z}=
\delta\varepsilon_{p}''/\varepsilon_{p0}''$. The range of values
of parameter ${y}= \Delta z/N$ which depends on ${x}=
(N_{p}+1)z/N$ and saturation factor ${f}=
\frac{1}{2}\tilde{\gamma}^{-2}\xi^{2} g(\tilde{\Omega})\tilde{I}$
is different in each domain (${N_{p0}}\equiv Q_{p0}/(\hbar
\omega_{mn} \gamma_{c})$ and ${N}\equiv Q_{T}/(\hbar \omega_{mn}
\gamma_{c})$ are the number of photons absorbed by the
nanoparticle during the time $\gamma_{c}^{-1}$ directly from the
light wave and the total one respectively.). However, parameter
$y$ is always negative in all domains of bistability.}
\end{center}
\label{fig5}
\end{figure}
\section{Conclusions}\label{c:con}
\begin{itemize}
    \item {Cascade energy transfer efficiency can rich as mach as
    several order of magnitude ($10^3 - 10^5$)}
    \item {Efficiency is drastically decreased at both large and  small distances between atom
    and nanoparticle surface}
    \item {For constant light frequency the efficiency as sharply as resonance depends      from the distance between atom and nanoparticle}
    \item {This sharp dependence can be used to determine the atom position near the
    surface}
    \item {Bistability may take place when the population difference and the relative growth of the image part of the particle dielectric function have the opposite signs.}
\end{itemize}

 The work was supported by RFBR, grant \# 02-02-17885.
\bibliographystyle{utphys_arXiv_hyperlinks}
\bibliography{Nikolaev}
\end{document}